\documentclass[12pt,english,reprint,twocolumn]{revtex4-1}
\usepackage[utf8]{inputenc}
\usepackage[a4paper]{geometry}
\usepackage{babel}
\usepackage{float}
\usepackage{amsmath}
\usepackage{amsthm}
\usepackage{graphicx}
\usepackage{esint}
\usepackage[unicode=true,
 bookmarks=true,bookmarksnumbered=false,bookmarksopen=false,
 breaklinks=false,pdfborder={0 0 0},pdfborderstyle={},backref=false,colorlinks=false]
 {hyperref}

\makeatletter

\usepackage{amsmath}
\usepackage{layout}
\usepackage[parfill]{parskip}
\usepackage[font=small,labelfont=bf]{caption}
\usepackage{eurosym}
\usepackage[utf8]{inputenc}
\usepackage{titlesec}

\titleformat*{\section}{\large\bfseries\raggedright}
\titleformat*{\subsection}{\normalsize\bfseries\raggedright}
\titlespacing*{\section}
{0pt}{5.5ex plus 1ex minus .2ex}{0.5ex plus .02ex}
\titlespacing*{\subsection}
{0pt}{5.5ex plus 1ex minus .2ex}{0.3ex plus .02ex}

\AtBeginDocument{

}

\makeatother

\begin{document}
\begin{abstract}

We demonstrate for a nonlinear photonic system that two highly asymmetric feedback delays can induce a variety of emergent patterns which are highly robust during the system's global evolution.
 Explicitly, two-dimensional chimeras and dissipative solitons become visible upon a space-time transformation.
 Switching between chimeras and dissipative solitons requires only adjusting two system parameters, demonstrating self-organization exclusively based on the system's dynamical properties.
 Experiments were performed using a tunable semiconductor laser's transmission through a Fabry-Perot resonator resulting in an Airy function as nonlinearity.
 Resulting dynamics were band-pass filtered and propagated along two feedback paths whose time delays differ by two orders of magnitude.
 An excellent agreement between experimental results and theoretical model given by modified Ikeda equations was achieved.

\end{abstract}

\title{Spatio-temporal complexity in dual delay nonlinear laser dynamics: chimeras and dissipative solitons}

\author{\textbf{D. Brunner$^{1,*}$, B. Penkovsky$^{1}$, R. Levchenko$^{2}$, E. Sch\"{o}ll$^{3}$, L. Larger$^{1}$, Y. Maistrenko$^{1,4}$}}

\affiliation{$^{1}$ FEMTO-ST Institute/Optics Department,CNRS \& Univ. Bourgogne Franche-Comt\'{e} CNRS, 15B avenue des Montboucons, 25030 Besan\c{c}on Cedex, France}
\affiliation{$^{2}$ Taras Shevchenko National University of Kyiv, Volodymyrska St. 60, 01030 Kyiv,Ukraine}
\affiliation{$^{3}$ Institut für Theoretische Physik, Technische Universit\"{a}t Berlin, Hardenbergstraße 36, 10623 Berlin, Germany}
\affiliation{$^{4}$ Institute of Mathematics and Center for Medical and Biotechnical Research, NAS of Ukraine, Tereschenkivska Str. 3, 01601 Kyiv, Ukraine \\
	$^{*}$ Corresponding author: daniel.brunner@femto-st.fr}

\maketitle

The complex dynamical properties of high-dimensional nonlinear systems continue to create new and fascinating phenomena.
 Already simple nonlinear equations or experimental systems are capable to produce dynamics ranging from highly coherent motion all the way to hyper-chaos \cite{Soriano2013}.
 Recently discovered chimera states even are combinations of both, chaotic and coherent motions within a symmetric network of identical elements \cite{Tinsley2012,Hagerstrom2012,Larger2015,Totz2017}; they have been recently reviewed \cite{PAN15,SCH16b}.
 This diversity stimulates not only continuous interest in the underlying principles, but has created a long-lasting output of novel applications.
 Nonlinear photonic systems have been identified as excellent substrates for highly coherent microwave oscillators \cite{Saleh2016}, chaos encryption \cite{Argyris2005}, neuromorphic processors \cite{Larger2012,Brunner2013,Vinkier2015,VanDerSande2017} as well as regenerative photonic memory \cite{Romeira2016}.

The complex motion of nonlinear dynamical systems often reveals its underlying structure in the form of geometric patterns.
 These are readily found in the spatio-temporal dynamics of two dimensional (2D) substrates.
 Prominent examples are dynamics found in Belousov-Zhabotinsky diffusion reactions \cite{Tinsley2012} liquid crystal displays \cite{ Hagerstrom2012,Arecchi2000,Verschueren2013} or broad area semiconductor lasers \cite{Barland2002,Ackermann2009,Descalzi2011}.
 Yet, these phenomena are not limited to spatially extended systems, and comparable dynamical objects exists in nonlinear dynamical systems coupled to delay \cite{Larger2013,Marconi2015,Yanchuk2017}.
 Such nonlinear delay systems are heavily exploited in photonic technology, i.e. in mode-locked fiber lasers.
 Since only recently nonlinear delay dynamics have been found to sustain stable laser chimera states \cite{Larger2013} as well as dissipative solitons (DS) \cite{Marconi2015}.

\begin{figure*}[t]
	\begin{centering}
		\includegraphics[width=\textwidth]{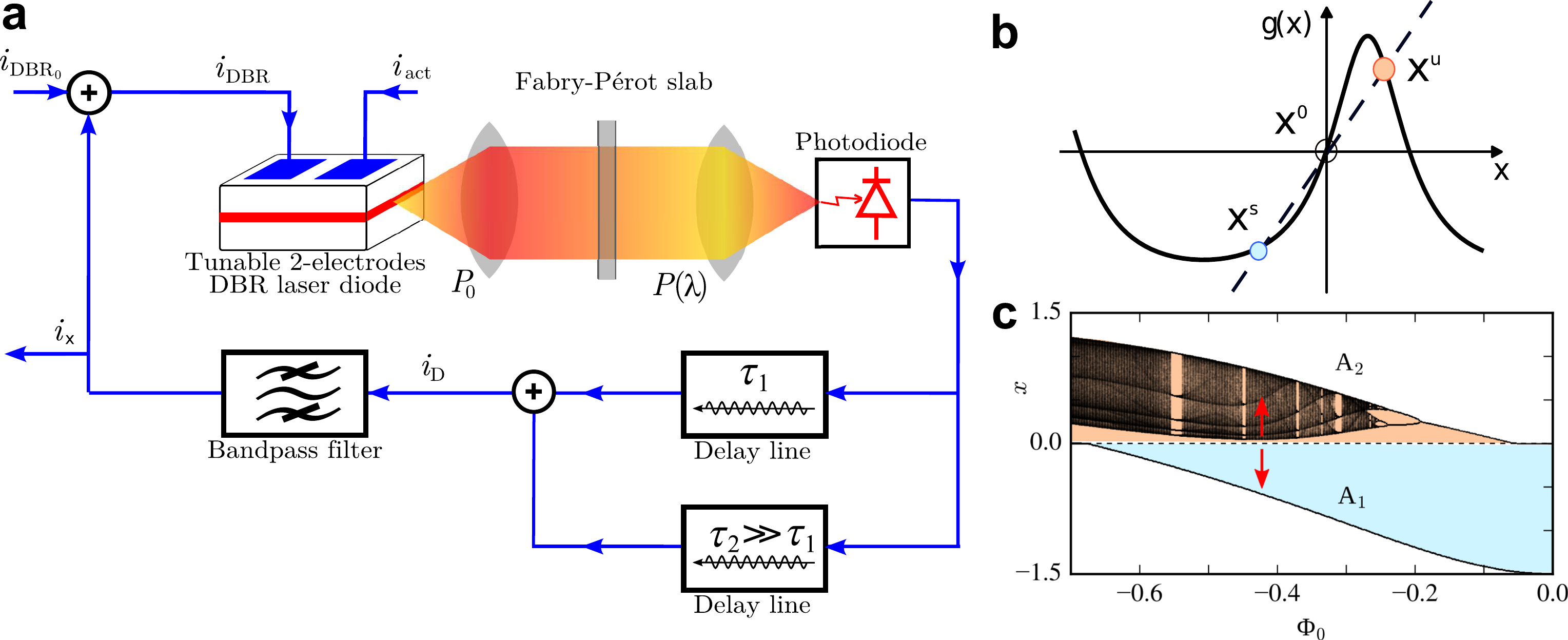}\caption{\textbf{| Experimental setup and its nonlinear delay dynamics.}
			(\text{a}): A semiconductor laser is subjected to two delayed optoelectronic feedbacks originating from a short $\tau_{1}$ and a long $\tau_{2}$$\gg\tau_{1}$ delay line.
			The delayed signals are bandpass filtered and injected into the DBR section of the laser diode, dynamically modulating its wavelength.
			(\textbf{b}): The optical emission is filtered by a Fabry-Perot filter, creating an Airy function as the system's nonlinear response.
			In a one dimensional (1D) map $g(x)$, the asymmetry of the Airy function results in multiple fixed points $x^s$ and $x^u$, identified by its intersection with the shown first bisector (dashed line).
			(\textbf{c}): The 1D map's bifurcation diagram, with resulting dynamics concentrated around the system's fixed points.
			Red arrows indicate attracting direction for positive or negative initial conditions, identifying the range of dynamics found on attractor A$_1$ (light blue) and A$_2$ (light red).
			A$_1$ corresponds to stable fixed point $x^s$, A$_2$ to an invariant attracting set born from unstable fixed point $x^u$.}
		\label{fig:experimental_setup}
		\par\end{centering}
\end{figure*}

Complex nonlinear dynamics found in delay systems relies on the finite propagation speed of the signal along the feedback path.
 In addition, the resulting propagation delay establishes a mapping between temporal and virtual space positions \cite{Giacomelli1996,Arecchi1992}.
 After a normalization by approximately the delay-time, consecutive sections of unity length can be stacked.
 The result are dynamics in one continuous space dimension with a second dimension representing integer time \cite{Giacomelli1996,Larger2015,PAN15,SCH16b}.
 While spatially extended systems typically possess two dimensions, delay systems have so far been mostly limited to a single virtual space dimension \cite{Yanchuk2017}.
 Here, we overcome this critical limitation and investigate pattern formation in a delay system coupled to two independent delays.
 A single photonic nonlinearity is coupled to feedback originating from two different feedback paths, where one delay exceeds the other by two orders of magnitude.

\begin{figure*}[t]
	\begin{centering}
		\includegraphics[width=0.9\textwidth]{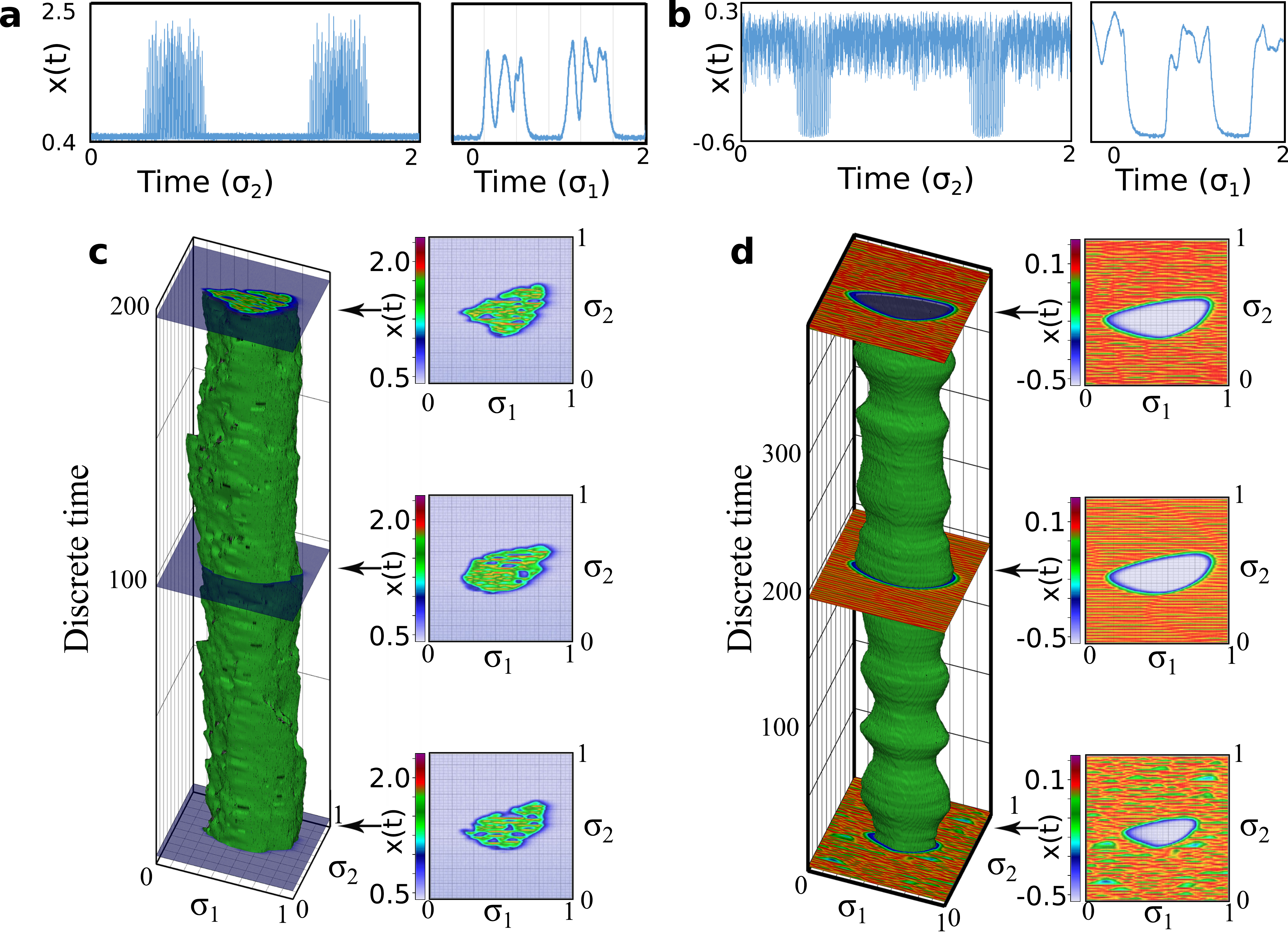}
		\caption{\textbf{| Experimental laser chimeras.}
		Asymptotic temporal waveforms of the chimeras during a long interval of the length $2\tau_{2}$ and a zoom focusing on the short delay interval of the length $2\tau_{1}$ are shown in panels (\textbf{a}) and (\textbf{b}).
		Corresponding space-time dynamics and respective snapshots of the chimeras in their 2D-virtual space are shown in (\textbf{c}) and (\textbf{d}).
		Temporal evolution along the vertical dimension is accompanied by slow, chaotic breathing.
		Parameters are $\beta \approx 1.4$, $\Phi_{0} = -0.6$, $\gamma = 0.5$, $m=33.5$ and $\beta \approx 0.7$, $\Phi_{0} = -0.05$, $\gamma = 0.5$, $m=33.5$ for panel \textbf{a,c} and \textbf{b,d}, respectively.}
	\label{fig:exp_chimeras}
	\par\end{centering}
\end{figure*}

We report on multiple types of chimeras and DS in a 2D virtual space consisting of a grid of coupled virtual photonic nonlinear oscillators \cite{Larger2015}.
 Assigning the evolution of the system's 2D space to a third dimension, chimeras and DS form free standing three dimensional (3D) structures.
 Chimeras create columns consisting either of a coherent steady state or of chaotic dynamics, while the excitable nature of the DS manifests itself in spatial chaos \cite{Coullet1987,Firth1988} due to a random spatial position \cite{Marconi2015,Romeira2016}.
 Extensive theoretical analysis accompanies our experimental investigations.
 Using delay differential equations (DDE) we confirm our findings via an excellent agreement between numerical simulations and experiment.
 Based on a reduced map equation, we identify the distribution of the dynamical variable during a particular state as the ordering mechanism of the different 2D dynamical objects, i.e., 2D chimeras and DS.
 
\section*{Results}

In Fig. \ref{fig:experimental_setup}a we schematically illustrate the experimental setup \cite{Larger2013}.
 A tunable semiconductor laser diode, emitting at $\sim$1550 nm, is biased at its gain section with a current of $i_{act}=$20 mA. 
 The laser's wavelength is controllable via a second electrode, which supplies current $i_{DBR}$ to the distributed Bragg-reflector (DBR) section \cite{Larger2015}.
 As illustrated, the DBR electrode receives current from two sources and $i_{DBR}=i_{DBR0} + i_{x}$.
 $i_{DBR0}$ is externally set to a constant value, while the second contribution corresponds to our delay system's dynamical variable $i_{x}$.
 The laser's optical intensity $P_{0}$ is detected after propagation through a Fabry-Perot filter, making the detected power $P(\lambda, t)$ a function of the laser's wavelength at present.
 Our system's nonlinearity therefore corresponds to
\begin{equation}
f(x)=\beta\left(1+m\cdot\sin^{2}\left[x+\Phi_{0}\right]\right)^{-1},\label{eq:airy}
\end{equation}
 where $f(x)$ is the Airy nonlinear function and $\beta$ is a linear amplification.
 In our case, the nonlinearity is created by a simple Farby-Perot resonator.
 The spectral width of the filter's transmission window is determined by $m$, which is a function of the resonator surfaces' reflectivity. 
 $\Phi_{0}$ depends on the static wavelength controlled by $i_{DBR0}$, and $x$ is the dynamical variable of our system proportional to the dynamic DBR current $i_{x}$.
 Physically, the detected signal is the photocurrent $i_D \propto f(x)$.

$i_D$ is consecutively divided and delayed along two delay lines with signal delays $\tau_{1}$ and $\tau_{2}$.
 Both delays were implemented using first-in first-out memory blocks of a field programmable gate array (FPGA), and we chose $\tau_{2} = 100 \times \tau_{1}$. 
 The FPGA recombines both delayed signals according to $i_D(t)=(1-\gamma)i_D(t-\tau_{1}) + \gamma i_D(t-\tau_{2})$, where $\gamma$ allows for relative weighting between the two delays.
 Finally, $i_D$ is bandpass filtered and scaled with feedback gain $\beta$, creating current $i_{x}$.
 The double delay loop is closed by injecting $i_{x}$ into the laser's DBR section.

In Fig. \ref{fig:experimental_setup}b we illustrate the nonlinear Airy function together with the first bisector.
 Multiple fixed points exist: one is located on a positive slope located in a flat plateau, the other on a steep negative slope close to the transmission maximum. 
 In order to better understand their dynamical properties, we evaluate the 1D map $f :x\mapsto f(x)$ against $\Phi_{0}$ and at $\beta = 1.6$ and $m = 4.7$.
 In Fig. \ref{fig:experimental_setup} (c) we show the resulting bifurcation diagram of $g(x) = f(x) - f(0)$, demonstrating the impact the asymmetry of $f(x)$ exerts upon the existence and stability of these fixed points.
 The small positive slope around $x^s$ results in the regular fixed-point attractor A$_{1}$, the steep negative slope around $x^u$ gives rise to a chaotic attractor A$_{2}$ via a period-doubling cascade.
 Dynamics around these fixed points create basins of attraction, which are separated by the unstable fixed point $x^{0}$.
 Within both basins, the system's dynamical variable covers ranges $U(A_{1}(\Phi_{0}))$ and $U(A_{2}(\Phi_{0}))$, indicated via blue and orange areas in Fig. \ref{fig:experimental_setup}c, respectively.
 Parameter $\Phi_{0}$ shifts function $f(x)$ along the horizontal axis and as such adjusts the ration between $U($A$_{1})$ and $U($A$_{2})$.
 Combined with the bandpass filter's integrating effect, this ratio imposes a global structure upon the dynamics.

Using scaling $\tilde{t} = \frac{t}{\tau_{1} + \rho_{1}}$, we link time $t$ to virtual space dimensions \cite{Larger2013,Larger2015} according to
\begin{align}
	\sigma_{1}(t) =& \tilde{t} - \lfloor \tilde{t} \rfloor \label{eq:sigma1} \\
	\sigma_{2}(t) =& \frac{\lfloor \tilde{t} \rfloor \tau_{1}}{\tau_{2} + \gamma_{2}} , \label{eq:sigma2}
\end{align} 
where $\lfloor \tilde{t} \rfloor$ is the floor of $\tilde{t}$, i.e., the greatest integer that is less than or equal to $\tilde{t}$.
 Using a time trace section of length  $\tau_{1} + \rho_{1}$ and $\tau_{2} + \rho_{2}$, respectively,  $\sigma_{1}(t)$ and $\sigma_{1}(t)$ create a 2D map resulting in a pseudo-space of unit size.
 Dividing the full time trace into non-overlapping sections of such length creates discrete temporal sequences of the 2D system's long-time evolution.
 The small constants $\rho_{1}$ and $\rho_{2}$ are linked to the response time of the oscillator $\varepsilon$ \cite{Giacomelli2012}.

\subsection*{2D Laser chimeras}

\begin{figure*}[t]
	\begin{centering}
		\includegraphics[width=0.95\textwidth]{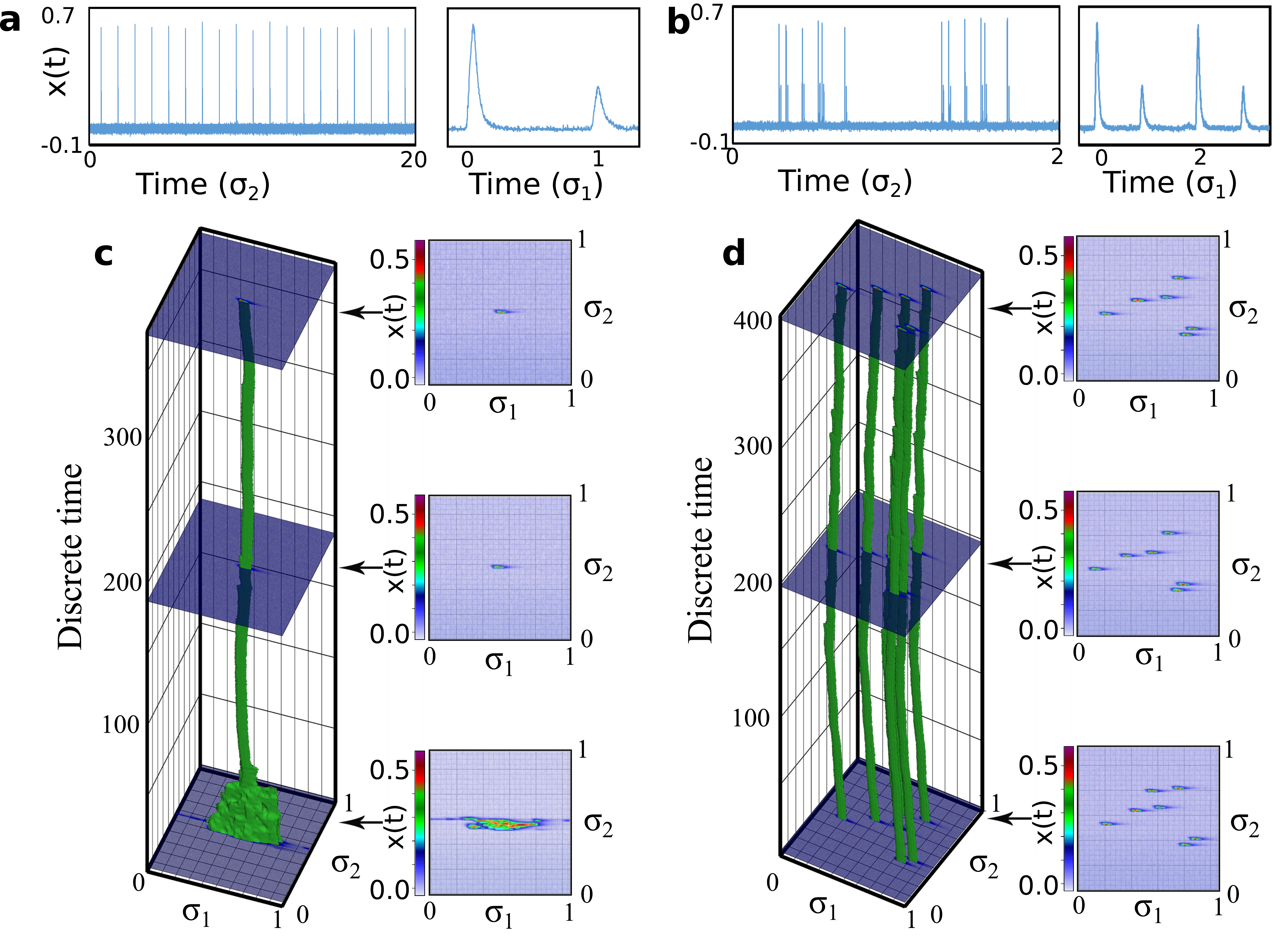}
		\par\end{centering}
	\caption{\textbf{| Dissipative solitons obtained experimentally.}
		Panel (\textbf{a}) shows the transition of chimera-like initial condition to a single DS, while in panel (\textbf{b}) we show multiple DS.
		Asymptotic temporal waveforms of the solitons at the long delay time intervals and a zoom at the short delay intervals are shown on the top of both panels.
		3D space-time plots and respective snapshots in 2D of the DS are shown below.
		As illustrated by the multiple DS state of panel \textbf{b}, DS are randomly located within the 2D space as they are induced by random initial conditions.
		Parameters were $\beta = 1.4$, $m=33$, $\gamma=0.75$ and $\Phi_0 = -0.6$.}
	\label{fig:exp_DS}
\end{figure*}

Figure \ref*{fig:exp_chimeras}a,b show the temporal evolution of two different types of chimeras found in our experiment.
 Parameters were $\beta\approx$ 1.4, $\Phi_{0}\approx$ -0.6 and $m=$ 33.
 In Fig. \ref*{fig:exp_chimeras}a, complex dynamics are broken into sections spaced by a steady state lasting $\sim\tau_{2}$.
 As revealed by the second, shorter time trace, dynamical motion inside a $\sim\tau_{2}$ window is further structured on temporal scales close to the short delay $\tau_{1}$.
 Generally speaking, the system predominantly resides in a steady state, regularly alternating with chaotic dynamics.
 The stable fixed point dynamics are located on attractor A$_1$, while chaotic motion resides on the chaotic attractor A$_2$.
 At the chosen $\Phi_{0}$ the stable fixed point's amplitude is less than the amplitude range covered by dynamics along the chaotic attractor.
 As a consequence, the majority of time the system resides close to its stable fixed point, from where it makes excursions through the chaotic attractor's complex phase space.
 Globally, the pattern formed by these alternations is iteratively stable after the recurrence of the long delay.

In order to reveal the globally stable character, the full dynamical motion can be better captured after transformation into the spatio-temporal dimensions according to Eqs. (\ref{eq:sigma1}) and (\ref{eq:sigma2}), with the resulting 3D representation shown in Fig. \ref{fig:exp_chimeras}c.
 Inside the ($\sigma_{1}, \sigma_{2}$)-plane, a region of chaotic dynamics is enclosed by the stable fixed point state.
 Along the third dimension we only show the boundary separating chaotic and stable motion.
 Showing the evolution of this state along the vertical discrete time dimension, and besides slight modifications, we demonstrate the long term persistence of this 2D structure.
 As shown in \cite{Larger2015,Yanchuk2017}, our 2D spatio-temporal system representation can be interpreted as a 2D network of nonlinear oscillators, each oscillator experiencing identical coupling and node parameters.
 Employing the analogy to such a network of Kuramoto phase oscillators, oscillators in the stable fixed point all share a common phase, while in the chaotic state no such uniform phase relationship exists.
 In our dual delay system such symmetry breaking can exclusively be attributed to dynamical properties; the deviations from a perfect coupling symmetry, which can hardly be avoided in a 2D substrate, can here be excluded.
 Combined with the temporal stability, this identifies our dynamical state as the first demonstration of a chimera state along two virtual space dimensions of a double delay system.

Upon changing parameters to $\beta\approx0.7$, $\Phi_{0}\approx$-0.05, the stable fixed point dynamics become surrounded by a sea of chaotic motion, as can be seen from data shown in Fig. \ref{fig:exp_chimeras}b.
 Similar to a harbor without physical walls inside a turbulent sea, this island of tranquility stably exists for longer than 350$\tau_{2}$.
 We therefore find that parameter $\Phi_{0}$ is an essential characteristic for the pattern formation in our system.
 Laser chimeras with incoherent core arise in the case when $U(A_{2})$ is larger, while a larger $U(A_{1})$ gives rise to the chimeras with a coherent core.
 Finally, for equal sizes of $U(A_{1})$ and $U(A_{2})$ we expect stripe like chimera states.
 Both reported chimera states were obtained for $\gamma=0.5$, for which the short and long delayed feedbacks have equal amplitude weights.
 We generally find that chimera states arise under such balanced amplitude scaling conditions for both delays.

\begin{figure}[h!]
	\begin{centering}
		\includegraphics[width=0.9\columnwidth]{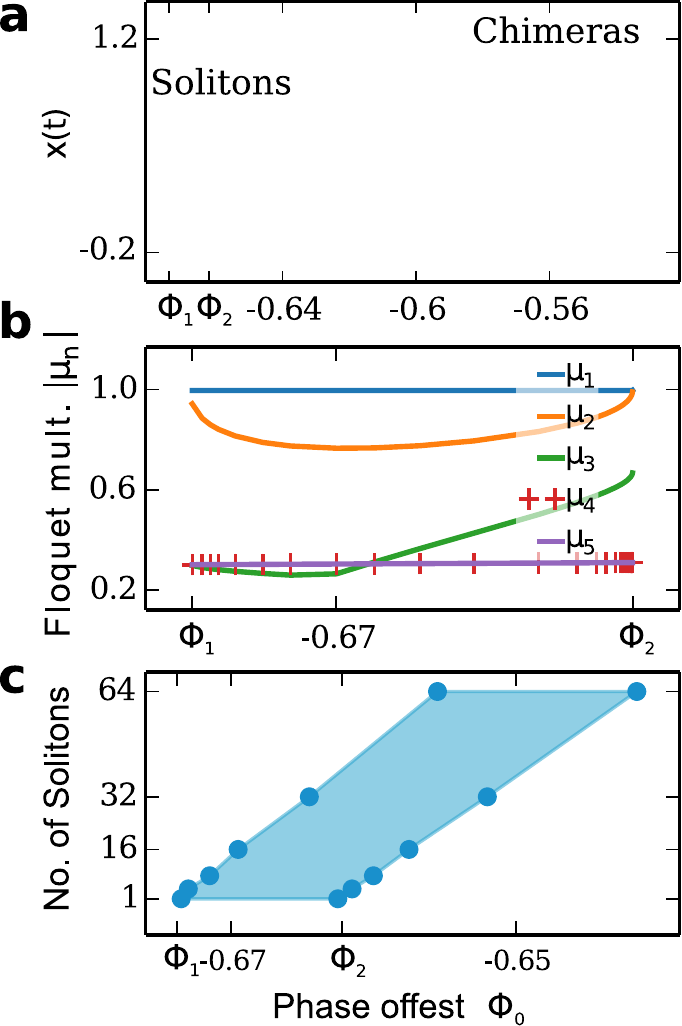}
		\par\end{centering}
	\caption{\textbf{| Numerical investigation.}
		Panel (a) shows the good agreement between the experiment and our numerical model for comparable parameters.
		Panel (b) shows the five largest Floquet-multipliers for the single DS state and their dependency on $\Phi_0$. 
		The bifurcation diagram of the DS is shown in panel (c).
		Increasing $\Phi_0$ enables the system to support more and more DS.
		Beyond a certain feedback strength, solutions containing few DS are damped.    
 	    Parameters fixed in all numerical simulations: $\beta=1.6$, $\gamma=0.83$, $m=50$, $\varepsilon= 10^{-2}$, $\delta = 9\cdot 10^{-3}$, $\tau_2 = 100 \tau_1$.
		}
	\label{fig:sim_DS}
\end{figure}

\subsection*{Dissipative solitons}

Upon further exploration of the system parameters, we find that the chaotic attractor A$_2$ can play a different, quite counter-intuitive role.
 For more negative phase values, $\Phi_{0}\approx-0.68$ and $\beta\approx1.5$, attractor A$_2$ transforms into a Cantor set while the limit cycle fixed point $x^{s}$ of A$_1$ exchanges stability with the origin $x^{0}$ in a transcritical bifurcation.
 As a result, the states available to our system are limited by the stable fixed point $x^0$ and a number of solutions arising under the influence of the attractor located in $U($A$_{2})$.
 Two characteristic examples of resulting dynamical states are illustrated in Fig. \ref{fig:exp_DS}a,b.
 Stable fixed point dynamics occupy the majority of the system's temporal evolution, interrupted by isolated spike-like structures.
 As we chose $\gamma = 0.75$, the long delay's impact significantly dominates over the short delay.
 Consequently, dynamics are strongly regular on the $\tau_{2}$-scale but not on scale $\sim\tau_{1}$.
 We find that dynamics can experience one, Fig. \ref{fig:exp_DS}a, or multiple spikes per delay $\tau_{2}$, Fig. \ref{fig:exp_DS}b.

Again, the global dynamical property can be better appreciated after the 2D transformation.
 The single spike of Fig. \ref{fig:exp_DS}a corresponds to a single dissipative soliton (DS) \cite{Marconi2015}.
 The temporal evolution along the discrete long-time axis, Fig. \ref{fig:exp_DS}c, reveals that this particular DS is born from chimera-like initial conditions which, in the course of time, transform into the asymptotic waveform of the DS.
 Data experiencing multiple spikes per $\sigma_{2}$ correspond to multiple DS structures, an example is shown in Fig. \ref{fig:exp_DS}d.
 Each of the shown DS structure was obtained by resetting the setup (blocking the laser beam), and multiple initializations of the system gave rise to different DS structures. 
 Most DS observed structures stably persisted over long time scales.

\section*{Theoretical analysis and discussion}

For numerical simulations we use a double-delayed Ikeda equation including the integral term $\delta\cdot\intop_{t_{0}}^{t}x(\xi)d\xi$.
 Physically, this integral term corresponds to adding a high-pass filter which is essential for stable two-dimensional patterns \cite{Larger2013}.
 The resulting equation is
\begin{equation}
\begin{aligned}
\varepsilon \frac{dx(t)}{dt}+x(t)+\delta\cdot\intop_{t_{0}}^{t}x(\xi)d\xi = \\
(1-\gamma)f\left(x_{\tau_{1}}\right)+\gamma f\left(x_{\tau_{2}}\right).\label{eq:chimera-2}
\end{aligned}
\end{equation}
Here, $x(t)$ is the dynamical variable, $f(x)$ the nonlinear transformation given by Eq. (\ref{eq:airy}), and $\varepsilon$ and $\delta$ are small time-scale parameters.
 The nonlinear transformation acts on both delayed feedback contributions $x_{\tau_{1}}=x(t-\tau_{1})$ and $x_{\tau_{2}}=x(t-\tau_{2})$, with the time delay ratio set to $\tau_{2}/\tau_{1}=100$.
 Relative amplitude scaling between the two feedback loops is provided by $\gamma$.
 We find an excellent agreement between our experiment and numerical simulations based on Eq. (\ref{eq:chimera-2}) at parameters comparable to the experiment.
 In the model, multiple yet strongly damped echoes appear on a $\tau_{1}$-scale.
 Due to the inherent noise, only three of such echoes can be recognized in the experiment.
 Here we would like to point out that in the case of DS we analyze numerical simulations based on a narrower Airy function with $m = 50$.
 Under these conditions DS are stable in a significantly larger region than for $m=33$ as used in the experiment.
 This allows for a more substantiated analysis of the underlying phenomena, which is our objective.
 The resulting insight is transferable to the experimental system.

In Fig. \ref{fig:sim_DS}a we show the bifurcation diagram scanning $\Phi_{0}$ at $m=$ 50, $\beta = 1.6$, $\varepsilon$= 0.01 and $\delta$= 0.009.
 One can clearly identify the regions where DS and chimera dynamics can be found.
 The DS region ($\Phi_{1} \leq \Phi_{0} \leq \Phi_{2}$) is located left to the discussed transcritical bifurcation, where the stable fixed point $x^{s}$ is replaced by $x^{0}$ ($\Phi_{0}<-0.595$).
 Chimeras are only supported by the system when its steady state corresponds to the fixed point $x^{s}$ ($\Phi_{0}>-0.595$).
 We then analyze the stability of a single DS in detail using the DDE-BIFTOOL package \cite{DDEBIFTOOL}.
 In Fig. \ref{fig:sim_DS}b, the modulus of the five largest Floquet multipliers are shown for different values of $\Phi_{0}$.
 The largest multiplier is real and always equal to 1, corresponding to the motion along the limit cycle. 
 All other multipliers lie inside the unit circle, guaranteeing stability of the limit cycle.
 At the ending points $\Phi_{1}$ and $\Phi_{2}$ of the DS interval, the second largest and exclusively real-valued Floquet-multiplier approaches 1 following a standard square root dependency.
 Solitons therefore disappear at these boundaries due to a fold limit cycle bifurcation: the stable limit cycle is approached by an unstable one,
 they coalesce and annihilate each other.
 We therefore conclude that in our system the single DS structure arises via a fold limit cycle bifurcation.
 Details of fold limit cycle bifurcations can be found in \cite{Kuznetsov1995} Ch.5.3 and \cite{Izhikevich2007} Ch.6.3.
  
Figure \ref{fig:sim_DS}c shows the DS's stability intervals containing up to 64 DSs
obtained via direct numerical simulations.
 The existence window for a single DS-solution excellently agrees with the Floquet-multipliers shown in Fig. \ref{fig:sim_DS}b.
 However, for a range in $\Phi_{0}$, numerous solutions with different number of DS are supported by the system.
 The width in $\Phi_0$ of this region remains almost constant.
 The different numbers of co-existing DS-states for one $\Phi_{0}$ corresponds to a large multistability, and our simulations confirm that comparable fold limit cycle bifurcation scenarios occur for these multi-DS structures \cite{Kuznetsov1995}.
 Such peculiar multistability resembles behavior in the neighborhood of a saddle-focus homoclinic orbit with one unstable direction along $x(t-\tau_2)$ and a focus-spiraling behavior close to the ($x(t), x(t-\tau_1)$)-plane.
 We have observed such a structure in our numerical simulations as well as in the experiment for as long as the finite signal to noise ratio permits.
 Cycle multistability is a well known behavior in the neighborhood of saddle-focus homoclinic trajectories \cite{Shilnikov2001}.
 Further analysis of this complex issue is, however, beyond the scope of this work.
 
In our simulations, multiple DSs were seeded using initial conditions based on the experimental data.
 Our simulations however show that the underlying fold limit cycle bifurcation scenario can be extended to arbitrary DS location in the virtual 2D phase space.
 We are therefore confident that more efficient spatial distributions exist and one could significantly increase the number of DS solutions, making such systems interesting for optical memory \cite{Romeira2016}.
 Finally, we would like to stress that in addition to the DS we have also confirmed the agreement between our developed numerical model and data recorded from the experiment with regards to the chimera states.
 When transformed into the space-time illustrations, the agreement between simulations and experiment data is excellent.

\section*{Conclusion}

In conclusion, we have demonstrated long living complex structures corresponding to chimeras and dissipative solitons in a highly asymmetric double delay system.
 After a transformation into a 2D pseudo-space, these dynamical states manifest themselves as free standing columns along the third dimension.
 It is the first time that this space/time analogy was shown in experiments on double delay systems.
 We anticipate that dual-delay systems and their related dynamical phenomena will represent a simple yet powerful tool for further investigations of complex self-organized motions in two dimensions.
 The perfect symmetry of networks implemented in double delay systems is a fundamental asset and presents a unique opportunity to compare two dimensional models and hardware systems.
 The capacity to generate high-dimensional, yet stable patterns might also open new possibilities to the currently very active explorations into neuromorphic computing using nonlinear systems.
 In this strongly emerging topic, during a learning phase complex patterns need to be generated and then stored e.g. for efficient and fast classification performed by a convolution operation at the read-out layer of a  Reservoir Computer \cite{Larger2012,Brunner2013}.

\section*{Acknowledgments}

The authors acknowledge the support of the Region Bourgogne Franche-Comt\'{e}.
 This work was supported by Labex ACTION program (contract ANR-11-LABEX-0001-0) and by Deutsche Forschungsgemeinschaft (DFG) in the framework of SFB 910.
 Last but not least, the authors thank Jan Sieber for fruitful discussions and his valuable assistance with the DDE-BIFTOOL package.

\end{document}